\documentclass[reprint,amsmath,amssymb,aps,]{revtex4-1}

\usepackage{ulem}
\usepackage{color}
\usepackage{amsmath}
\usepackage{graphicx}
\usepackage{dcolumn}
\usepackage{bm}

\begin{document}
\preprint{APS/123-QED}

\title{A Hybrid Quantum Memory Enabled Network at Room Temperature}

\author{Xiao-Ling Pang,$^{1,2}$ Ai-Lin Yang,$^{1,2}$ Jian-Peng Dou,$^{1,2}$ Hang Li,$^{1,2}$ Chao-Ni Zhang,$^{1,2}$ Eilon Poem,$^{3,4}$ Dylan J. Saunders,$^{3}$ Hao Tang,$^{1,2}$ Joshua Nunn,$^{3,5}$ Ian A. Walmsley,$^{3}$ }
\author{Xian-Min Jin$^{1,2,6}$}
\email{xianmin.jin@sjtu.edu.cn}

\affiliation{$^1$School of Physics and Astronomy, Shanghai Jiao Tong University, Shanghai 200240, China}
\affiliation{$^2$Synergetic Innovation Center of Quantum Information and Quantum Physics, University of Science and Technology of China, Hefei, Anhui 230026, China}
\affiliation{$^3$Clarendon Laboratory, University of Oxford, Parks Road, Oxford OX1 3PU, UK}
\affiliation{$^4$Department of Physics of Complex Systems, Weizmann Institute of Science, Rehovot 7610001, Israel}
\affiliation{$^5$Centre for Photonics and Photonic Materials, Department of Physics, University of Bath, Claverton Down, Bath BA2 7AY, UK}
\affiliation{$^6$Institute for Quantum Science and Engineering and Department of Physics, Southern University of Science and Technology, Shenzhen 518055, China}

\pacs{Valid PACS appear here}
\maketitle 

{\bf Teaser: Building quantum network with room-temperature atomic and all-optical memories for scalable quantum information processing.}

{\bf Quantum memory capable of storage and retrieval of flying photons on demand is crucial for developing quantum information technologies. However, the devices needed for long-distance links are quite different from those envisioned for local processing. Here, we present the first hybrid quantum memory enabled network by demonstrating the interconnection and simultaneous operation of two types of quantum memory: an atomic-ensemble-based memory and an all-optical loop memory. The former generates and stores single atomic excitations that can then be converted to single photons; and the latter maps incoming photons in and out on demand, at room-temperature and with a broad acceptance bandwidth. Interfacing these two types of quantum memories, we observe a well-preserved quantum cross-correlation, reaching a value of 22, and a violation of the Cauchy-Schwarz inequality up to 549 standard deviations. Furthermore, we demonstrate the creation and storage of a fully operable heralded photon chain state that can achieve memory-built-in combining, swapping, splitting, tuning and chopping single photons in a chain temporally. Such a quantum network allows atomic excitations to be generated, stored, and converted to broadband photons, which are then transferred to the next node, stored, and faithfully retrieved, all at high speed and in a programmable fashion.}

\section*{Introduction}
\noindent A quantum network consisting of a large-scale distribution of quantum nodes and interconnecting channels remains an overarching goal for quantum information science. Such a network could be used for quantum-enhanced technologies which promise to outperform classical systems in the fields of communication, computing and metrology \cite{Jeremy_nphoton_2009}. Unfortunately, quantum channels suffer from exponential loss and high latency \cite{Xianmin_nphoton_2010,Gisin_nphoton_2007}. In addition, the probabilistic generation of quantum states limits the scale of quantum systems \cite{Ladd_nature_2010,Walter_nphys_2012}. Quantum memories capable of storing quantum states \cite{Lvovsky_nphoton_2009} permit quantum communication over long distances with quantum repeaters \cite{Briegel_PRL_1998,DuanLM_nature_2001}, and avoid the non-deterministic nature of quantum state generation by synchronizing stochastic events \cite{Nunn_PRL_2013,Paul_optica_2015,Eggleton_nCommun_2015,Paul_optica_2017,Paul_arXiv_2018}. 

Various works have been motivated on quantum memory theory and relevant physical implementations, such as optical delay lines and cavities \cite{Avramopoulos_OL_1993,Pittman_PRA_2002,Pittman_PRA_2002_cavity,Leung_PRA_2006}, electromagnetically induced transparency (EIT) \cite{Kuzmich_nature_2005,Eisaman_nature_2005,Xianmin_nphoton_2011}, the DLCZ protocol \cite{DuanLM_nature_2001,Kuzmich_nature_2003,Chrapkiewicz_PRL_2017}, photon-echo quantum memory \cite{Hosseini_nCommun_2011,Liao_PRL_2014}, off-resonant Faraday interaction \cite{Julsgaard_nature_2004}, Raman memory \cite{Reim_PRL_2012,DingDS_nphoton_2015}, and ATS memory \cite{Saglamyurek_nphoton_2018}. In order to make a quantum memory device practical for scalable and high-speed quantum information processing, the key requirements that have to be satisfied include broad acceptance bandwidth, high efficiency, long lifetime, low noise level, and the capability of running at room temperature.

However, for the last 20 years, it has been proven very challenging to meet all these requirements simultaneously. Considerable efforts have been made in pushing the bandwidth from the kHz and MHz regime to the GHz regime, and temperature from near absolute zero to room temperature \cite{Kuzmich_nature_2003,Julsgaard_nature_2004,Kuzmich_nature_2005,Eisaman_nature_2005,Xianmin_nphoton_2011,Hosseini_nCommun_2011,Reim_PRL_2012,DingDS_nphoton_2015,Chrapkiewicz_PRL_2017}. At room temperature, EIT and near off-resonant Raman memories suffer from fluorescence noise which is impossible to be filtered out spectrally \cite{PanJW_PRA_2007}. As for far off-resonance Raman memory, while it eliminates fluorescence noise, a new intrinsic noise related to spontaneous Raman scattering emerges \cite{Michelberger_NJP_2015}. 

\begin{figure*}
	\centering
	\includegraphics[width=1.6\columnwidth]{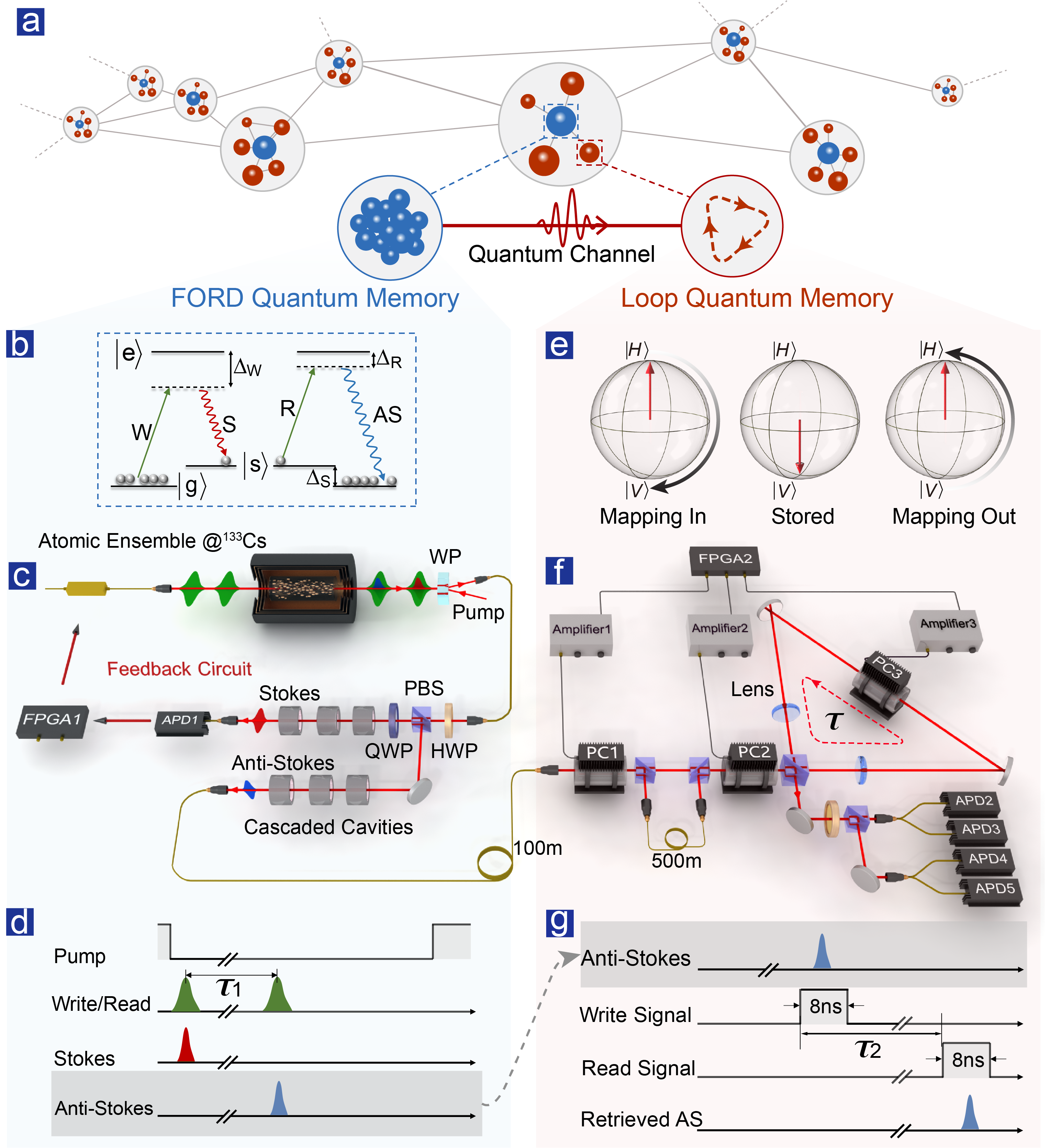}\\
	\caption{\textbf{A schematic diagram and experimental setup for a hybrid quantum memory enabled network.
		} \textbf{a.} A quantum network consists of two different-functional nodes and their interconnections. \textbf{b.} Write and read processes of FORD quantum memory. Solid lines represent three-level $\Lambda$-type configuration of atoms, in which states $\left| {\rm{g}} \right\rangle $ and $\left| {\rm{s}} \right\rangle $ are hyperfine ground states of cesium atoms ($\Delta_s$ = 9.2\,GHz); state $\left| {\rm{e}} \right\rangle $ is excited state; dash lines represent broad virtual energy levels induced by the write and read pulses. \textbf{c.} Setup of FORD quantum memory. WP: Wollaston prism. PBS: polarization beam splitter QWP: quarter wave plate, HWP: half wave plate. \textbf{d.} Time sequences of FORD quantum memory. \textbf{e.} Polarization switching in the mapping in-and-out processes shown in Bloch spheres. \textbf{f.} Setup of Loop quantum memory. The Pockels cell in loop is controlled by write and read electrical signals from two channels of a field-programmable gate array (FPGA) module. A 500-meter-long fiber is introduced to coordinate with the Loop memory as another switching path against photon loss. Four avalanche photodiodes (APDs) are used to detect photons in a chain with small time interval. PC: Pockels cells. \textbf{g.} Time sequences of Loop memory. The time interval $\tau_2$ between write and read signals can be any positive integral multiples of one cycle period $\tau$.}
	\label{f1}
\end{figure*}

Recently, we have realized a broadband and room-temperature Far Off-Resonance Duan-Lukin-Cirac-Zoller (FORD) quantum memory, capable of operating with a high fidelity in the quantum regime \cite{Dou_Commun_2017}. Now we are pursuing a quantum memory that is broadband, room-temperature, and more importantly, compatible with the FORD quantum memory for further storing Stokes/anti-Stokes photons (i.e. mapping in) and retrieving them (i.e. mapping out) for controlled durations, without any additional noise. Progress towards such a memory has been made by using excited states of atoms \cite{Kaczmarek_arXiv_2017,Finkelstein2018} albeit with a limited lifetime. 

Alternatively, loop architecture has shown the capability to trap and control photons by employing ultra-low-loss optical elements, which can avoid introducing any additional noise \cite{Kuzmich_nature_2005,ChenZB_PRA_2007,Michelberger_NJP_2015,DingDS_nphoton_2015}. The idea of loop architecture can be traced back to 1993 \cite{Avramopoulos_OL_1993}, and though being seemingly straightforward, it is actually very challenging to realize in practice. Continuous efforts have been made to improve its performance and practicability \cite{Pittman_PRA_2002,Pittman_PRA_2002_cavity,Leung_PRA_2006,Paul_optica_2015,Paul_arXiv_2018,Paul_optica_2017,Christine_science_2011}, and recently, extremely low loss has been achieved, enabling high-efficiency single- \cite{Paul_optica_2015,Paul_arXiv_2018} and multi-photon generation \cite{Paul_optica_2017}. A Loop quantum memory, consisting of an all-optical storage loop and controllable polarization switches, can map broadband flying single photons in and out at room temperature, which thus complements the FORD quantum memory.

Here, we propose and experimentally demonstrate a broadband, room-temperature and hybrid quantum memory enabled network. The hybrid network combines a broadband FORD memory with a compatible all-optical Loop memory. The network generates non-classical optical states in a built-in and time-controllable fashion in a warm atomic ensemble, and traps these states in an all-optical loop (see Figure 1a). This hybrid quantum memory system is capable of combining, swapping, splitting and chopping single photons in a chain temporally, as well as finely tuning the time of each individual photon. Some of these tasks have been realized in the frequency domain for coherent light \cite{Hosseini_nature_2009}, but have never been achieved in the time domain or for genuine quantum light. After traversing the two memory devices successively, the quantum states are observed to be high-fidelity Fock states. We measure the cross-correlation and find that it exceeds the classical Cauchy-Schwarz bound \cite{Clauser_PRD_1974} by more than 500 standard deviations, establishing that our memories would be capable of violating a Bell inequality \cite{BaoXiaohui_nphoton_2012}, enabling the detection of non-classical behavior and that they could be deployed in quantum applications.

\begin{figure*}
	\includegraphics[width=1.95\columnwidth]{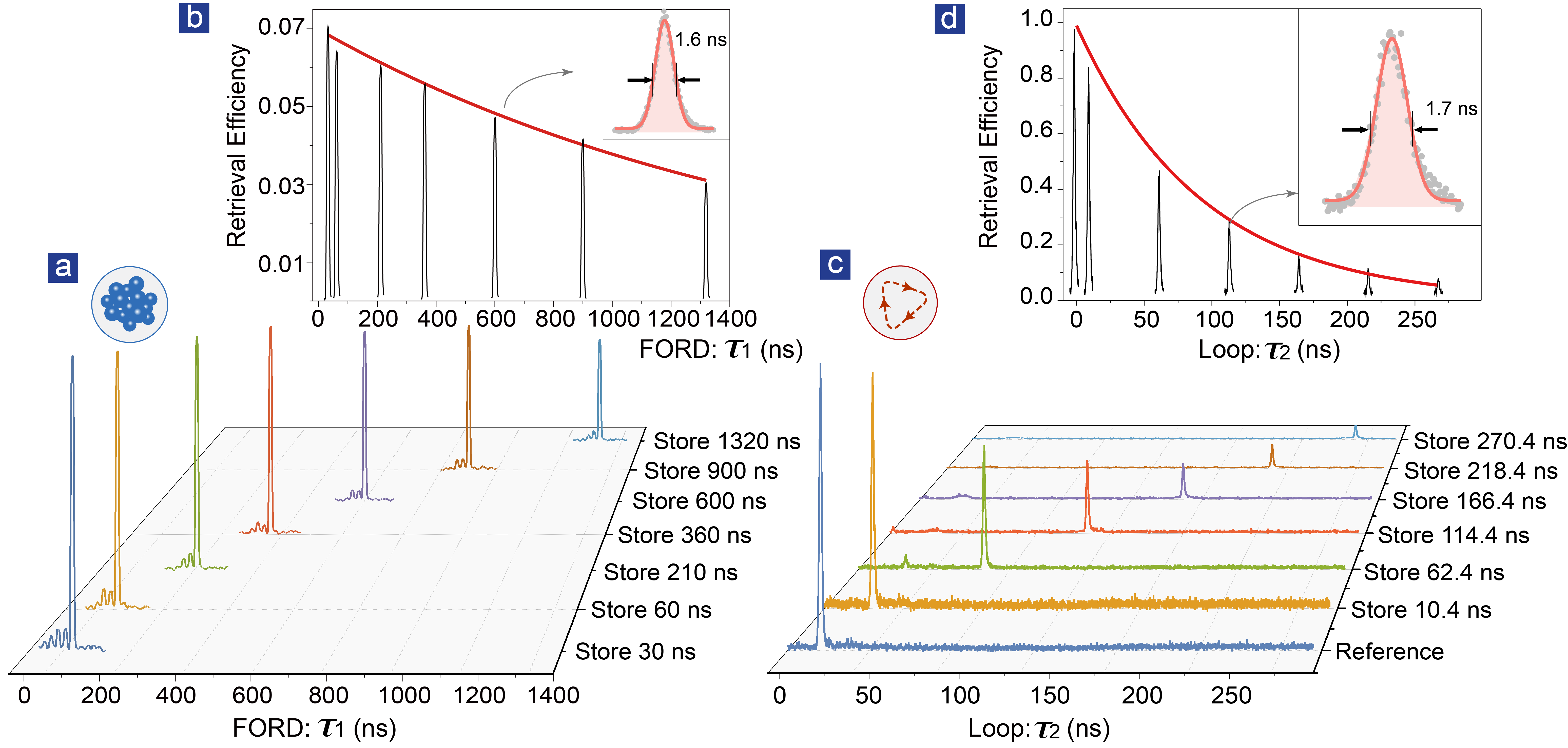}\\
	\caption{\textbf{Retrieval efficiency of FORD and Loop memory}. \textbf{a.} The retrieved distribution of anti-Stokes photons triggered by Stokes photons as a function of $\tau_1$. \textbf{c.} The retrieved distribution of injected photons as a function of $\tau_2$, with the circulation period $\tau$ being 10.4\,ns. \textbf{b} and \textbf{d.} The red solid lines fit the measured retrieval efficiency of the FORD and Loop memories. Inset figures show the pulses at $\tau_1=600$\,ns and $\tau_2=114.4$\,ns with Gaussian fits, giving the duration of the retrieved broadband pulses. Comparing with the duration of 1.6\,ns  measured before the Loop memory, the pulse shape is well preserved. The transmission per roundtrip is 90\% (two Pockels cell are utilized in the loop with the same clock rate of 50\,kHz and rise time of 5\,ns, and the transmittance for each Pockels cell is 97\%).}
	\label{f2}
\end{figure*}

\section*{Results}
The schematic view of the FORD protocol is shown in Figure 1(b,c,d). A strong write/read pulse is incident on an atomic ensemble, creating a Stokes photon via spontaneous Raman scattering, which heralds the successful creation of an atomic excitation \cite{DuanLM_nature_2001}. After a programmable delay ${\tau _1}$, the corresponding anti-Stokes photon is retrieved, where the retrieval efficiency in principle can also reach 100\% \cite{Reim_PRL_2012}. The strong and broadband write/read pulses undergo a much larger detuning than that is usually applied in the standard DLCZ protocol, which makes the acceptance bandwidth very broad for Stokes/anti-Stokes photons. The strong write pulse with a detuning ${\Delta _{\rm{W}}}$ creates a broadband Stokes photon, while the atomic ensemble with a single created excitation becomes \cite{DuanLM_nature_2001,Nicolas_RMP_2011}
\begin{equation}\label{eq1}
\left| \psi  \right\rangle {\rm{ = }}\frac{1}{{\sqrt N }}\sum\limits_{j = 1}^N {{e^{\mathrm{i}\left( {{{\vec k}_W} - {{\vec k}_S}} \right) \cdot {{\vec r}_j}}}} \left| {{g_1}{g_2}...{s_j}...{g_N}} \right\rangle \
\end{equation}
where $N$ is the number of atoms, ${\vec k_W}$ (${\vec k_S}$) is the wave vector of the write (Stokes) pulse, and ${\vec r_j}$ is the coordinate of the ${j}$th atom which is excited to state $\left| {\rm{s}} \right\rangle $. After a programmable time delay ${\tau _1}$, a read pulse with a detuning ${\Delta _{\rm{R}}}$ transforms the collective excitation state into a broadband anti-Stokes photon. The measured bandwidth is about 500\,MHz, calculated by a convolution-based approach \cite{Dou_Commun_2017}.

Connected by an optical fiber, flying anti-Stokes photons are transmitted to the remote all-optical Loop memory. As is shown in Figure 1 (e,f,g), once the anti-Stokes photon is mapped in the loop, high-speed polarization switches (Pockels cells) are activated. The first control pulse applied to the Pockels cell in the loop switches the photon to be vertically polarized. The photon then cycles counterclockwise for a controllable time $\tau_2$, until the second control pulse applied to the Pockels cell maps it out by switching its polarization back to horizantal. Then the retrieved anti-Stokes photon is collected by a fiber coupler, and a standard Hanbury Brown-Twiss experiment can be conducted to study photon statistics with a 50:50 fiber beam splitter and avalanche photodiodes. Remarkably, the bandwidth of our all-optical Loop memory is much higher than tens of THz, and is only limited by the response bandwidth of linear optical elements, which means the Loop memory matches the bandwidth of the FORD memory.

The retrieved distributions of anti-Stokes photons triggered by Stokes photons after the FORD memory are illustrated in Figure 2 (a,b). We attribute the slight decline of retrieval efficiency to be loss induced by atomic motion. The retrieved distribution of injected photons after the Loop memory are illustrated in Figure 2 (c,d). The retrieval efficiency of the Loop memory can be obtained by comparing the counts of the mapped-out anti-Stokes photons with the values before being mapped into the loop, which reaches 90\% for short storage times. The decrease mainly results from photon loss due to multiple reflection and dispersion from the optical elements in the loop. The red fitted curve in Figure 2d implies that our current devices (two Pockels cells are being upgraded to a single fast and low-loss one) enable light storage for around 10 cycles in the loop. Insert figures show that the pulse shapes associated with the broadband nature of the released anti-Stokes photons are well preserved.

The main limitation of the FORD memory's retrieval efficiency here is the lack of another independent laser generation system. In our current experimental setup, since the frequencies of write and read pulses are the same, the detuning of read process is much smaller than that of write process. As a result, the read process may happen immediately after an excitation is created, leading to the limited retrieval efficiency. The problem can be solved by preparing the write and read pulses independently with different lasers. Therefore, the read effect in the write process can be suppressed greatly, or can even be ignored, resulting in much higher retrieval efficiency. In addition, other possible solutions include increasing the pulse energy \cite{Reim_PRL_2012}), optimizing the pulse shape \cite{Novikova_PRA_2008}, harnessing cavity enhancement \cite{Saunders_PRL_2016} and using a double-pass geometry \cite{Guo_OL_2017}.

\subsection{Quantum correlations in a hybrid quantum memory network}

In order to evaluate the ability of our memories to preserve quantum correlations, we measure the second-order correlation function $g_{{\rm{S\text{-}AS}}}^{(2)}\left( {{\tau_1,\tau_2}} \right)$ between the retrieved anti-Stokes mode and the Stokes mode. The results are presented in Figure 3 as functions of storage time ${\tau _1}$ and ${\tau _2}$. The decrease of $g_{{\rm{S\text{-}AS}}}^{(2)}\left( {\tau _1},{\tau _2=31.2{\rm{\,ns}}} \right)$ of the FORD memory (see Figure 3a) mainly comes from atomic motion \cite{Zhao_nPhys_2009} and background noise, while the slight drop of $g_{{\rm{S\text{-}AS}}}^{(2)}\left( {\tau_1=30{\rm{\,ns}},{\tau _2}} \right)$ of the Loop memory (see Figure 3b) comes from the decrease of the count rate as the efficiency drops. At the initial values of storage time, the nonclassicality can be revealed clearly by a violation of Cauchy-Schwarz inequality \cite{Clauser_PRD_1974}
\begin{equation}\label{eq2}
{\left( {g_{{\rm{S\text{-}AS}}}^{\left( 2 \right)}} \right)^2} \le g_{{\rm{S\text{-}S}}}^{\left( 2 \right)} \cdot g_{{\rm{AS\text{-}AS}}}^{\left( 2 \right)}
\end{equation}
by up to 549 standard deviations (the measured cross- and auto- correlations are $g_{{\rm{S\text{-}AS}}}^{\left( 2 \right)} = 22.63 \pm 0.93$, $g_{{\rm{AS\text{-}AS}}}^{\left( 2 \right)} = 1.57 \pm 0.50$, $g_{{\rm{S \text{-} S}}}^{\left( 2 \right)} = 2.11 \pm 0.25$ ). The storage time at which the cross-correlation drops to $1/e$ of its maximum value \cite{Zhao_nPhys_2009,Xianmin_nphoton_2011} can be viewed as the maximum storage time over which the memories preserve quantum coherence and quantum correlations. This lifetime, for the FORD memory, is 1.45\,${\rm{\mu s}}$, while for the loop memory this lifetime is 1.22\,${\rm{\mu s}}$. Note that the retrieval efficiency of the loop memory decays more quickly in time than the FORD memory, but the loop memory is also free of any noise processes such as fluorescence and four-wave mixing, which explains why it is capable of preserving quantum correlations for almost as long as the FORD memory. Indeed for a short storage time of 10.4\,ns, there is no measurable degradation in the $g_{{\rm{S\text{-}AS}}}^{\left( 2 \right)}$ cross correlation between the input and retrieved fields: for the input field we find $g_{{\rm{S\text{-}AS}}}^{\left( 2 \right)} = 21.1 \pm 0.8$ and for the retrieved field we measure $g_{{\rm{S\text{-}AS}}}^{\left( 2 \right)} = 21.5 \pm 1.3$.

\begin{figure*}
	\centering
	\includegraphics[width=1.74\columnwidth]{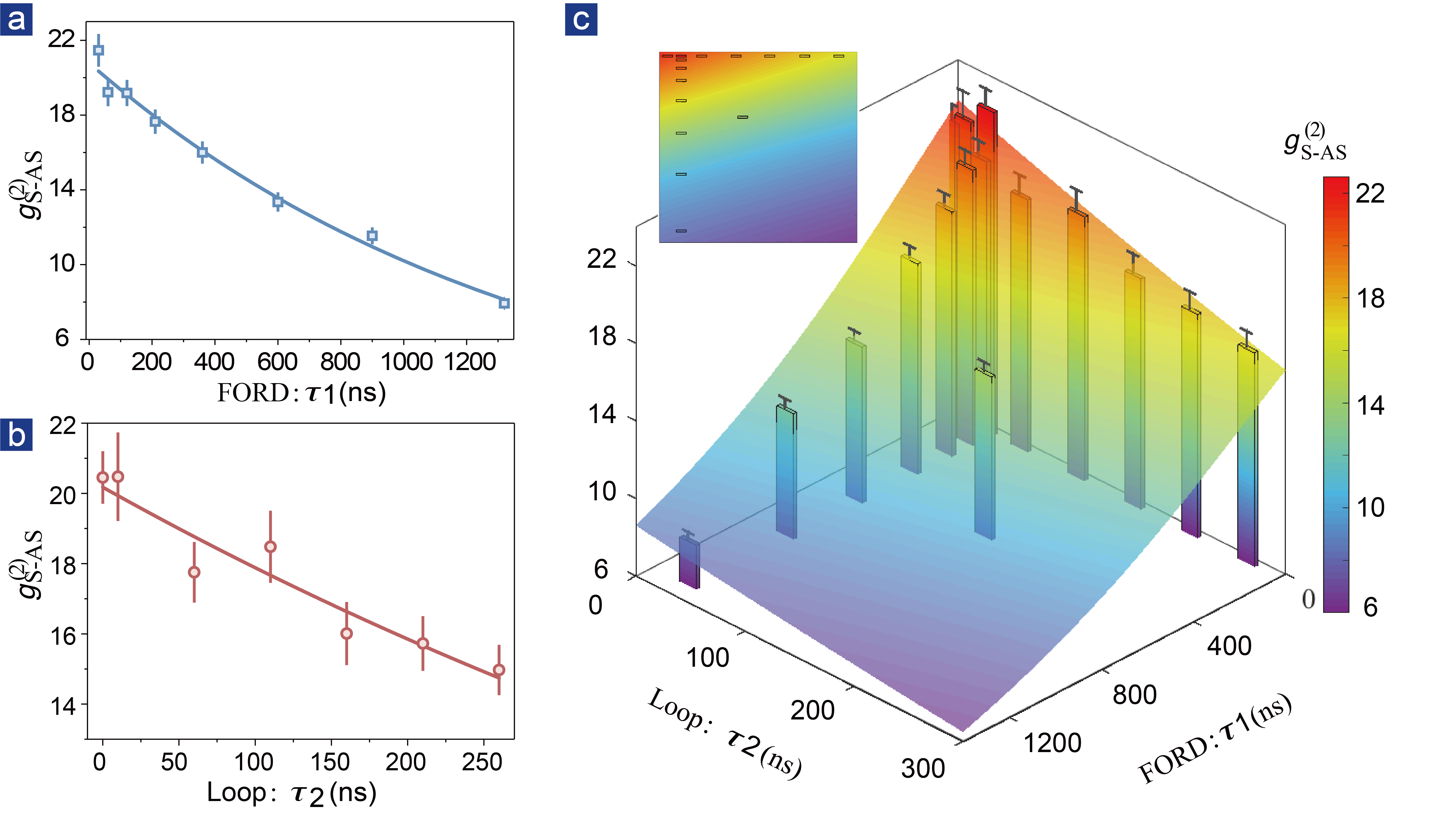}\\
	\caption{\textbf{Measured cross-correlation as a function of storage time.}  \textbf{a.} The cross-correlation $g_{{\rm{S\text{-}AS}}}^{\left( 2 \right)}\left( {\tau _1},{\tau_2=31.2{\rm{\,ns}}}\right)$ is fitted with form $g_{{\rm{S\text{-}AS}}}^{\left( 2 \right)}\left( {{\tau _1}} \right) = 1 + C/\left( {1 + A{\tau _1} + B{\tau _1}^2} \right)$, where the quadratic term describes atomic motion, and the linear term comes from background noise.  \textbf{b.} The cross-correlation $g_{{\rm{S\text{-}AS}}}^{\left( 2 \right)}\left( {\tau_1=30{\rm{\,ns}},{\tau _2}} \right)$ is fitted with form $g_{{\rm{S\text{-}AS}}}^{\left( 2 \right)}\left( {{\tau _2}} \right) = A{e^{ - B{\tau _2}}}$, which follows the exponential decay of the memory efficiency. The data are obtained under the condition of a write/read beam waist being 214\,$\mu$m, and the detuning $\Delta_R$ = 4\,GHz. Error bars are derived assuming Poissonian statistics of the individual photocounts. \textbf{c.} Joint cross-correlation of the hybrid quantum memory enabled network as a function of storage times. The surface is well described by the product of the two decay functions. The inset provides a top view of the measured points.}
	\label{f3}
\end{figure*}

\begin{figure*}
	\centering
	\includegraphics[width=1.9\columnwidth]{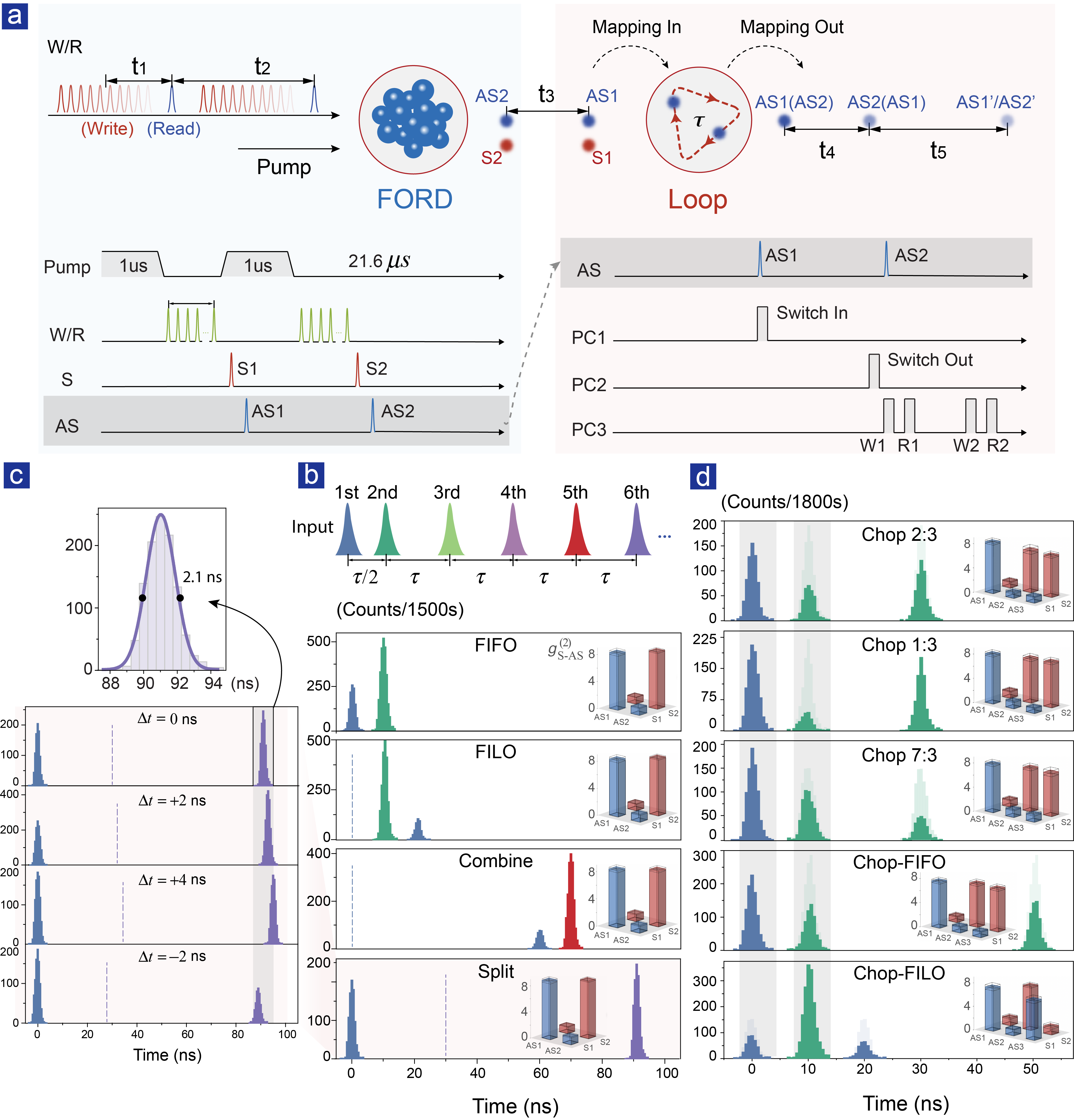}\\
	\caption{\textbf{Experimental creation and storage of a fully operable heralded photon chain state.}  \textbf{a.} Schematic diagram of creating a two-photon chain with tunable interval, and mapping the chain in and out to various temporal modes. $t_1$ is the storage time of FORD memory; $t_2$ is the time interval between two heralded anti-Stokes photons (AS1 and AS2); $t_3$ is the interval between AS1 and AS2 after a 500-meter fiber and two Pockels cells; $t_4$ represents the interval between AS1(AS2) and the first part of AS2(AS1); and $t_5$ is the interval between two chopped parts. The circulation period $\tau$ here is 20.3\,ns, where the transmission rate of the Pockels cell in the loop is 95\% with a clock rate of 30\,MHz and a rise time of 5\,ns. \textbf{b.} Various temporal modes of anti-Stokes photons are marked in different colors. Heralded photons can be mapped out of the Loop memory to arbitrary temporal modes defined. The transmission rate of anti-Stokes photons between two memories is around 12.4\%. FIFO: First In First Out; FILO: First In Last Out. \textbf{c.} Fine tuning of the mapped-out photons with a step of 2\,ns. The inset portrays the shape of the mapped-out anti-Stokes photons. \textbf{d.} Demonstration of chopping a single photon in a chain, where the probability is continuously tunable by adjusting half-wave voltage of the Pockels cell in the loop. The pulse shadows represent the sum of AS1(AS2) and AS1'(AS2'). Inserted bar graphs are measured cross-correlation values. More detailed values from $t_1$ to $t_5$ and cross-correlations can be found in Supplementary Materials.}
	\label{f4}
\end{figure*}

In a hybrid network suitable for quantum applications (Figure 1a), it should be possible to generate photons and to transfer them individually between quantum memories in a programmable way. Here we demonstrate one aspect of this functionality, by sequentially writing and then reading photons with intermediate storage times in both memories. As an example, we choose the point shown in the middle of Figure 3c, with $\tau_1 = 480\,ns$, $\tau_2 = 122.4\,ns$. The $g_{{\rm{S\text{-}AS}}}^{\left( 2 \right)}$ is found to be $14.43 \pm 0.47$, which is in good agreement with the prediction of 14.47 found using a simple model in which the best fit decay curves presented in Figure 3 (a,b) are multiplied together. In a large area, including the point (480{\rm{\,ns}},122.4{\rm{\,ns}}), the cross-correlations in such a hybrid quantum memory enabled network all exceeds the key boundary of 6. This, as suggested in \cite{BaoXiaohui_nphoton_2012}, means that the correlations between the Stokes and anti-Stokes photons are strong enough, in principle, to violate a Bell inequality.

\subsection{Creation and storage of a fully operable heralded photon chain state}

To better demonstrate the unique features of our hybrid quantum memory system, we further perform the storage of a fully operable heralded photon chain state. The hybrid quantum memory system is capable of combining, swapping, splitting and chopping single photons in a chain temporally, as well as finely-tuning the timing of each individual photon. To realize this, we add a feedback loop to repeat the spontaneous Raman process in the FORD memory until a Stokes photon is triggered successfully. Such a repeat-until-success scheme enabled by quantum memory increases the excitation rate up to 10 times. Meanwhile, the lifetime of the FORD memory is prolonged from 1.44\,$\mu$s to 2.24\,$\mu$s, by optimizing the spatial mode of the addressing light, in order to cover the feedback period which is around 1\,$\mu$s. A 500-meter-long fiber is introduced as a fixed delay line to coordinate with the Loop memory, which can mitigate the lifetime difference between the FORD and Loop quantum memories, allowing versatile two-memory coordination. Moreover, in order to store more temporal modes in a single loop, the circulation period is prolonged from 10.4\,ns to 20.3\,ns. Time sequences are shown as Figure 4a. More experimental details can be found in Supplementary Materials.

As is shown in Figure 4b, defining several temporal modes in different colors, two flying anti-Stokes photons in arbitrary input temporal modes can be mapped to any desired output temporal modes, presenting functions as: first in first out (FIFO), first in last out (FILO), combining and splitting two single photons. Furthermore, the time interval between two heralded anti-Stokes photons can be finely tuned by the delay of write pulses in the FORD memory, coordinating with the Loop memory in the temporal control of photons. In particular, as Figure 4c shows, the temporal mode of anti-Stokes photons can be finely tuned with a step of 2\,ns. Apart from temporal mode shifting, each of the single photons in the chain can be chopped into two temporal modes with tunable probability, shown in Figure 4d. The observed cross-correlation between two heralded anti-Stokes photons after two memories reaches 8.8, leading to a strong contrast with the measured cross-correlation of noise (1.0), and exceeding the bound of 6 needed to achieve a Bell violation. More measured cross-correlation values are shown in the Supplementary Materials.

\section*{Discussion}
The technical requirements for a quantum repeater are quite different from what are needed for local synchronisation. For a local photonic processor, entangling operations need to be attempted at a high rate, in order to build up a cluster state or similar, as a resource for quantum computing. Therefore, a memory must be able to store an incident optical mode that is already entangled with a wider system. The system clock rate should be high so that a large entangled state can be built, whereas latency is low. On the other hand, for a quantum repeater, only two parties are connected, so it is sufficient to distribute entanglement over a chain of linear segments. Bandwidth is important but not critical, whereas high signal latencies mean that a system with the potential for long storage times is valuable.

In addition, future photonic quantum processors could be securely connected over distance, provided that the primitive network elements are compatible. While we believe that the FORD quantum memory can be a building block for scalable quantum technologies, a network also requires memories that can store incoming photons, the functionality for which is provided by the Loop memory. Here we have shown that a Loop quantum memory for arbitrary input states, with high operating bandwidth but limited intrinsic lifetime, can behave as a node constituting a local photonic processor, and a FORD quantum memory, with long lifetime but technically more complex atomic-ensemble system, can serve as a main node linking other local photonic processors for constructing large-scale quantum networks.

In summary, we have proposed and experimentally demonstrated a broadband, room-temperature and hybrid quantum memory enabled network. The demonstrated building block consists of two different-functional quantum nodes: an atomic-ensemble-based memory capable of generating and storing quantum states, and an all-optical memory for mapping incoming photons in and out, at room-temperature and with high bandwidth. The two types of quantum memory are complementary in functionality, and are compatible with each other, both running at room-temperature and with broad bandwidth.

Our demonstration may provide a route towards scalable quantum information processing \cite{Ladd_nature_2010,Walter_nphys_2012,Lvovsky_nphoton_2009,Briegel_PRL_1998}, which requires the capability to create, store and distribute quantum states on-demand across a large numbers of nodes through interconnecting channels \cite{Kimble_nature_2008}. The performance of two types of quantum memories can be further improved in the future \cite{Balabas2010,Paul_optica_2017, Paul_arXiv_2018}, especially the ultra-low loss of loop architecture has been achieved experimentally. The currently achievable lifetime of the two quantum memories is at the microsecond level. However, the enabled hybrid quantum network would already be applicable in building quantum computers and quantum simulators locally.

\section*{Materials and Methods}
\textbf{Experimental details of FORD quantum memory:} We employ cesium atoms $^{133}$Cs to achieve a large optical depth due to its higher saturated vapor pressure compared with other alkali atoms. The 75-mm-long cesium cell with 10 Torr Ne buffer gas is placed into a magnetic shield, and is heated up to 61$^{\circ}$C to obtain an optical depth of about 5000. A horizontally polarized write/read laser beam creates Stokes and anti-Stokes photons with a programmable time lag. In our FORD memory, write and read pulses are co-propagating for maximizing the spin wave lifetime \cite{Zhao_nPhys_2009}. It has been shown that Stokes (anti-Stokes) photons are collinear with write (read) pulses \cite{DuanLM_PRA_2002}, but with orthogonal polarization \cite{Nunn_Ph.D_2008}, so we use a Wollaston prism to separate the retrieved photons from the strong addressing light. Meanwhile, since Stokes and anti-Stokes phtons are also co-propagating under the phase-matching condition, six home-built cavities arranged in a double pass configuration are employed to split Stokes and anti-Stokes photons. The transmission rate of each cavity reaches around $90\%$ while the extinction rate of each cavity against noise is up to 500:1.

\textbf{Experimental details of storing a heralded photon chain state.} We generate two heralded anti-Stokes (AS1 and AS2) photons successively in the FORD quantum memory in a built-in and time-controllable fashion. To maintain the high nonclassicality, the pump, read and write pulses are applied each time before an anti-Stokes photon is retrieved. Limited by the response rate of our Pockels cells, the operation period is 21.6\,$\mu$s, during which two anti-Stokes photons are created. After transmitting through the cavity filters, the first photon AS1 is selected to enter a 500-meter-long fiber with additional 2.5\,$\mu$s delay, mitigating the fixed time difference caused by the long state preparation period for the two anti-Stokes photons. The second photon AS2 does not pass through the fiber.

Since the circulation period of the Loop memory is 20.3\,ns and the rise/fall time of our Pockels cells is 5\,ns, a two-photon chain state can be stored in the Loop memory, and each of the photons can be addressed independently in the time domain. In our experiment, we demonstrated several functions that can be realized in such a hybrid quantum network. FIFO: the sequence of two mapped-out photons is conserved. FILO: the sequence of two mapped-out photons is reversed, by storing AS1 for more time than AS2. Combining: the two anti-Stokes photons with a large gap can be stored successively and retrieved as a two-photon chain state. Splitting: the gap between two anti-Stokes photons is increased by storing AS2 for more time than AS1. Chopping: each single photon in a chain can be mapped out with a tunable probability, and into different temporal modes on-demand. Furthermore, the fine-tuning of single photons can be realized by adjusting the time delay of the read process in FORD memory. The nonclassicality under all these situations is verified by the measured cross-correlation values. More details can be found in Supplementary Materials III.

\section*{Acknowledgments}
The authors thank Myungshik Kim and Jian-Wei Pan for helpful discussions. \textbf{Funding:} This research is supported by the National Key R\&D Program of China (2017YFA0303700), the National Natural Science Foundation of China (61734005, 11761141014, 11690033), the Science and Technology Commission of Shanghai Municipality (15QA1402200, 16JC1400405, 17JC1400403), the Shanghai Municipal Education Commission (16SG09, 2017-01-07-00-02-E00049). X.-M.J. acknowledges support from the National Young 1000 Talents Plan. \textbf{Author contributions:} X.-M.J. and I.A.W. conceived and supervised the project. X.-L.P. and X.-M.J. designed the experiment. X.-L.P., J.-P.D. and A.-L.Y. developed the laser control system. X.-L.P., A.-L.Y., J.-P.D., H.L., C.-N.Z.,E.P., D.J.S., H.T., J.N. and X.-M.J. performed the experiment and analyzed the data. X.-L.P. and X.-M.J. wrote the paper with input from all the other authors. \textbf{Competing interests:} The authors declare that they have no competing interests. \textbf{Data and materials availability:} All data needed to evaluate the conclusions in the paper are present in the paper and the Supplementary Materials. Additional data available from authors upon request.

\clearpage
\section*{\large Supplementary Materials}
\subsection*{I. Improved lifetime of FORD memory}
In order to further increase the memory lifetime to allow feedback-enhanced preparation of heralded single photon chain state, we optimize the spatial mode of read/write pulses in FORD quantum memory. The lifetime of FORD memory is defined as the value of cross-correlation dropping to $1/e$. Since the decrease of cross-correlation between Stokes and anti-Stokes photons is mainly caused by atomic motion, we enlarged the beam waist of write/read pulses from 214\,$\mu$m to 385\,$\mu$m. Therefore, it takes more time for excited atoms to run out of interaction region, which brings longer lifetime of FORD quantum memory from 1.44\,$\mu$s to 2.24\,$\mu$s, as is shown in Figure 5. The results are fitted by $g_{{\rm{S\text{-}AS}}}^{\left( 2 \right)}\left( {{\tau _1}} \right) = 1 + C/\left( {1 + A{\tau _1} + B{\tau _1}^2} \right)$, where quadratic term comes from atomic motion (Ref. 43), and linear term comes from background noise. In addition, the spatial mode of control light is also optimized as Gaussian beams, which can be seen form the insets of Figure 5.

\begin{figure*}
	\centering
	\includegraphics[width=2\columnwidth]{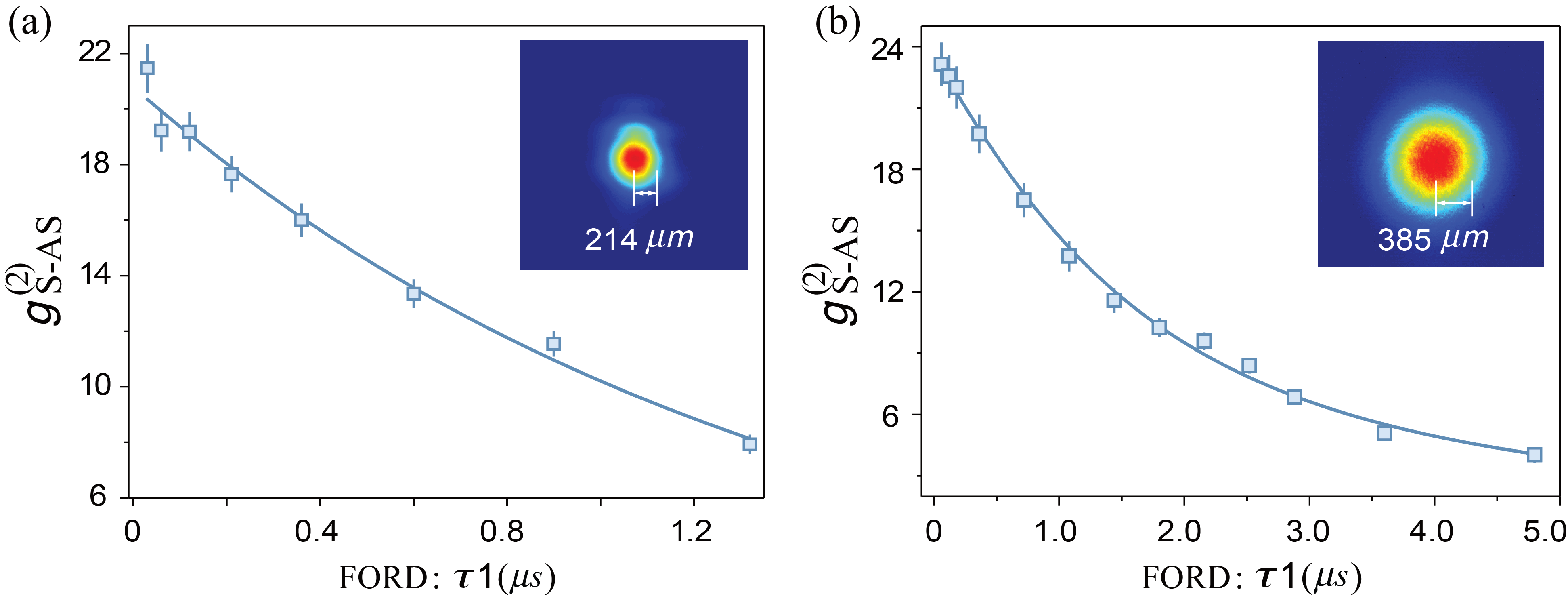}\\
	\caption{\textbf{Measured cross-correlation as a function of storage time.} \textbf{(a)} Cross-correlation functions measured by a write/read beam waist of 214\,$\mu$m. \textbf{(b)}  Cross-correlation functions measured by a write/read beam waist of 385\,$\mu$m. The insets are the measured cross-sections of control lights in the middle of atoms, and the values of beam waist are calculated by Gaussian fitting.}
	\label{f1}
\end{figure*}

\begin{figure*}
	\centering
	\includegraphics[width=2\columnwidth]{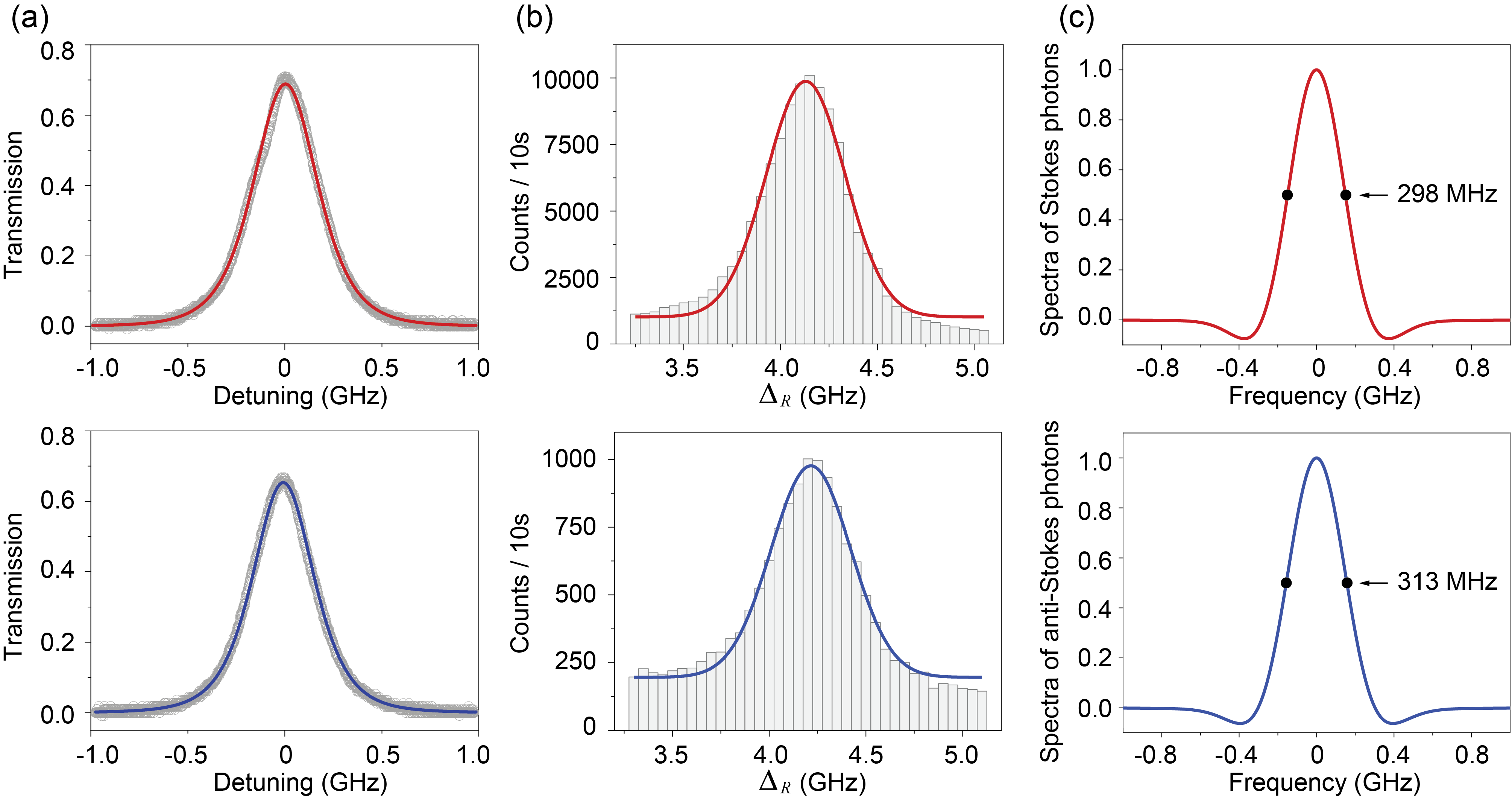}\\
	\caption{\textbf{Bandwidth measurement of Stokes (red curves above) and anti-Stokes photons (blue curves below).} \textbf{(a)} Total transmission spectra of cascaded cavity filters for Stokes and anti-Stokes photons. \textbf{(b)} Counts of transmitted Stokes and anti-Stokes photons obtained by scanning the detuning of write/read pulses. \textbf{(c)} Calculated frequency spectra by convolution theorem and Fourier transform. The marked bandwidth values represent for full-width at half-maximum.}
	\label{f2}
\end{figure*} 


\subsection*{II. Improved bandwidth-match between signal photons and cavity filters}
Since the polarizations of generated Stokes photons and anti-Stokes photons are the same, in our system, two sets of home-made cascaded Fabry-P{\'{e}}rot interferometers serve as frequency filters, resonant with Stokes photons and anti-Stokes photons respectively. In the experimental results shown in Figure 2 and 3 in the main text, the bandwidths of Stokes and anti-Stokes photons are around 500\,MHz. Later, we optimized the bandwidth precisely to around 300\,MHz to match our cavity filters. The newly obtained results of storing the photon chain state shown in Figure 4 are measured under the condition of the storage bandwidth being around 300\,MHz. Here we introduce the bandwidth measurement method that we have developed.

We firstly obtain the total transmission spectra of the cavities to be 396\,MHz for Stokes filters and 380\,MHz for anti-Stokes filters, shown in Figure 6(a). In order to match the bandwidth of cavity filters better, the bandwidth of FORD is optimized by using wider write/read pulses in the time domain (from 2\,ns to 4\,ns). According to the convolution-based approach (Ref. 31), scanning the detuning of write/read pulses, we record the distributions of Stokes and anti-Stokes photons, as shown in Figure 6(b). And eventually, the spectra of Stokes and anti-Stoke photons can be calculated by utilizing convolution theorem and Fourier transform. As is shown in Figure 6(c), the bandwidths of Stokes photons and anti-Stokes photons are around 300\,MHz.


\subsection*{III. Experimental scheme of operating a photon chain state}

Here we perform a new experiment of storing a fully operable heralded photon chain state, capable of combining, swapping, splitting and chopping single photons in a chain temporally, as well as finely tuning the time of each individual photon inside quantum memory. A heralded flying photon chain state is created via spontaneous Raman process of FORD memory, and is combined with the first heralded anti-Stokes photon (AS1) and the second heralded anti-Stokes photon (AS2). Then the photon chain can be mapped into the Loop memory, and after a programmable storage time, finally photons in the chain can be mapped out into various temporal modes independently. The number of photons that a loop can store depends on its circulation period and switching response rate. In our system, since the rise/fall time of Pockels cells is 5\,ns, we prolonged the circulation period of our loop from 10.4\,ns to 20.3\,ns, so that we can store a two-photon chain state with a 10\,ns interval between photons in the loop. Furthermore, since the interval between two photons is quite small, we upgrade the speed of polarization control system from 50\,kHz to 30\,MHz, making the operating of single photons in chain possible.

\begin{figure*}
	\centering
	\includegraphics[width=2\columnwidth]{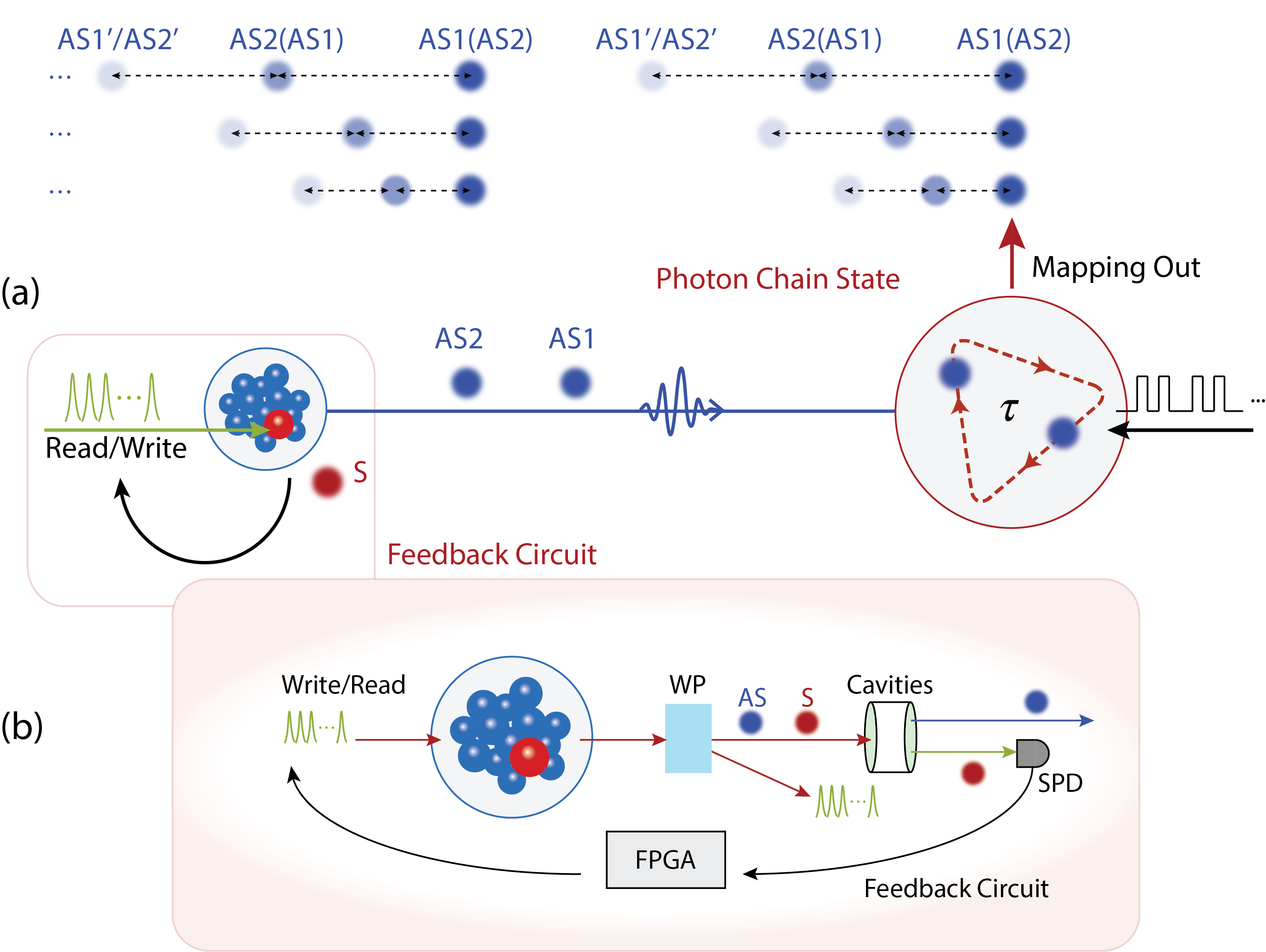}\\
	\caption{\textbf{Schematic diagram of creating and storing a photon chain state.} \textbf{(a)} The FORD memory generates two heralded anti-Stokes photons (AS1 and AS2) successively. Then photons in the chain can be mapped into the Loop memory via optical switches, and then be mapped out to designed temporal modes independently. \textbf{(b)} Detailed information of the feedback circuit. A Field Programmable Gate Array (FPGA) is used to generate write pulses ten times each period, until a Stokes photon is successfully detected. Wollaston prism (WP) and cavities serve as polarization and frequency filters respectively.}
	\label{f3}
\end{figure*}

Interesting things are that the interval between two photons in a chain can be finely tuned by FORD memory, and either of the photons can be mapped in and out of the Loop memory, with independent programmable storage time. Furthermore, each of the single photons in the chain can be chopped into two independent temporal modes with tunable probability, leading to two different output modes AS1(AS2) and AS1'(AS2'), of one single photon. The schematic diagram is shown in Figure 7. The observed cross-correlation between S1(S2) and mapped out AS1/AS1' (AS2/AS2') reaches 8.8, making a strong contrast with the measured cross-correlation of noise (1.0), and well exceeds 6, which ensures the violation of Bell's inequality. Measured values of cross-correlation functions are shown in Figure 8, and Table I. 

\begin{figure*}
	\centering
	\includegraphics[width=2\columnwidth]{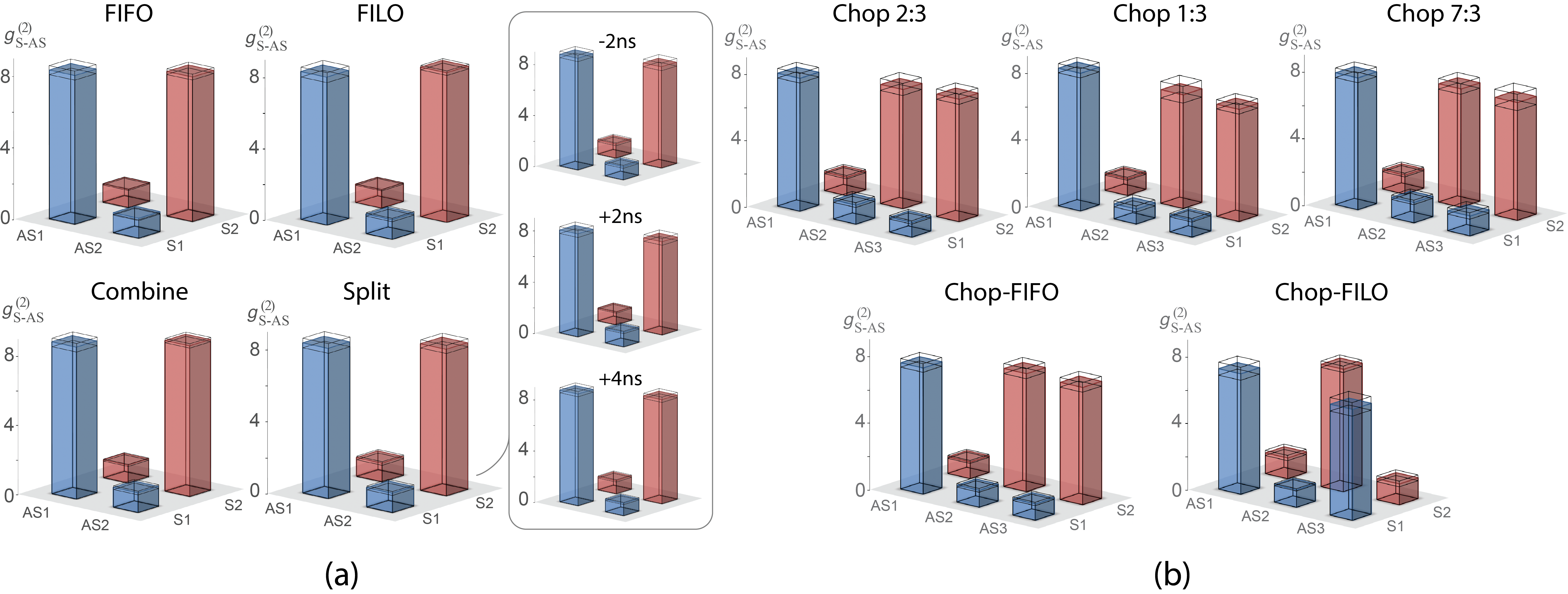}\\
	\caption{\textbf{Measured cross-correlation values for photon chain states.} \textbf{(a)} Cross-correlation values for various operations on AS1 and AS2: First In First Out (FIFO), First In Last Out (FILO), Combining and Splitting. Three inserts in the right are the cross-correlations of fine tuning situations. \textbf{(b)} Cross-correlation values for various operations on AS1, AS2 and AS2', under various chopping probabilities.}
	\label{f4}
\end{figure*}

\begin{table*}[ht]
	\centering	
	\caption{\textbf{Measured cross-correlation functions under various operations.}}	
	\label{tab1}
	\begin{tabular}{p{2.1cm}<{\centering} p{1.2cm}<{\centering} p{1.2cm}<{\centering}p{1.1cm}<{\centering}p{1.1cm}<{\centering}p{1.1cm}<{\centering}p{1.4cm}<{\centering}p{1.4cm}<{\centering}p{1.4cm}<{\centering}p{1.4cm}<{\centering}p{1.4cm}<{\centering}p{1.4cm}<{\centering} }
		\hline\noalign{\vskip 0.6mm}
		\hline\noalign{\vskip 0.2cm}		
		 	 & $t_1$(ns) & $t_2$(ns) & $t_3$(ns) &$t_4$(ns) & $t_5$(ns) & $g^{(2)}_{S1-AS1}$ & $g^{(2)}_{S2-AS2}$ & $g^{(2)}_{S1-AS2}$ & $g^{(2)}_{S2-AS1}$ & $g^{(2)}_{S2-AS3}$ & $g^{(2)}_{S1-AS3}$	 \\[0.10cm]		
		\hline\noalign{\vskip 0.11cm}		
		\textbf{FIFO}	 & 565.5 &	 2488.5 & 10 & 10 & - & $8.3\pm0.3$ & $8.1\pm0.2$ & $1.0\pm0.1$ & $1.1\pm0.1$ & - & - \\[0.10cm]		
		\hline\noalign{\vskip 0.11cm}		
		\textbf{FILO} & 565.5 &	 2488.5 & 10 & -10 & - & $8.3\pm0.3$ & $8.4\pm0.1$ & $1.0\pm0.0$ & $1.1\pm0.1$ & - & - \\[0.10cm]		
		\hline\noalign{\vskip 0.6mm}
		\textbf{Combine} & 565.5 & 2548.5 & 70 & 10 & - & $8.8\pm0.3$ & $8.7\pm0.2$ & $1.0\pm0.1$ & $1.1\pm0.1$ & - & - \\[0.10cm]	
		\hline\noalign{\vskip 0.6mm}
		\textbf{Split} & 565.5 &	2508.5 & 30 & 90 & - & $8.0\pm0.3$ & $8.1\pm0.1$ & $1.0\pm0.0$ & $1.1\pm0.1$ & - & - \\[0.10cm]		
		\hline\noalign{\vskip 0.6mm}
		\textbf{(+2\,ns)} & 565.5 & 2510.5 & 32 & 92 & - & $8.0\pm0.3$ & $7.2\pm0.3$ & $1.1\pm0.1$ & $1.0\pm0.1$ & - & - \\[0.10cm]	
		\hline\noalign{\vskip 0.6mm}
		\textbf{(-2\,ns)} & 565.5 & 2506.5 & 28 & 88 & - & $8.7\pm0.3$ & $7.9\pm0.3$ & $1.0\pm0.1$ & $1.1\pm0.1$ & - & - \\[0.10cm]		
		\hline\noalign{\vskip 0.6mm}
		\textbf{(+4\,ns)} & 565.5 & 2512.5 & 34 & 94 & - & $8.7\pm0.2$ & $8.1\pm0.2$ & $0.9\pm0.1$ & $1.0\pm0.1$ & - & - \\[0.10cm]	
		\hline\noalign{\vskip 0.6mm}
		\textbf{Chop 1:3} & 565.5 & 2488.5 & 10 & 10 & 20 & $8.2\pm0.3$ & $6.8\pm0.5$ & $0.9\pm0.2$ & $1.0\pm0.1$ & $6.7\pm0.3$ & $1.0\pm0.1$ \\[0.10cm]	
		\hline\noalign{\vskip 0.6mm}
		\textbf{Chop 2:3} & 565.5 & 2488.5 & 10 & 10 & 20 & $8.0\pm0.3$ & $7.2\pm0.4$ & $1.1\pm0.1$ & $1.1\pm0.1$ & $7.4\pm0.3$ & $0.9\pm0.1$ \\[0.10cm]		
		\hline\noalign{\vskip 0.6mm}
		\textbf{Chop 7:3} & 565.5 & 2488.5 & 10 & 10 & 20 & $7.9\pm0.3$ & $7.1\pm0.3$ & $1.2\pm0.1$ & $1.1\pm0.1$ & $7.0\pm0.5$ & $1.2\pm0.2$ \\[0.10cm]		
		\hline\noalign{\vskip 0.6mm}		
		\textbf{Chop-FIFO} & 565.5 & 2488.5 & 10 & 10 & 40 & $7.5\pm0.3$ & $7.1\pm0.3$ & $1.0\pm0.1$ & $1.0\pm0.1$ & $7.0\pm0.3$ & $1.0\pm0.1$ \\[0.10cm]
		\hline\noalign{\vskip 0.6mm}
		\textbf{Chop-FILO} & 565.5 & 2488.5 & 10 & 10 & 10 & $7.2\pm0.4$ & $7.4\pm0.2$ & $1.0\pm0.1$ & $1.2\pm0.2$ & $1.2\pm0.2$ & $6.6\pm0.4$ \\[0.10cm]	
		\hline\noalign{\vskip 0.6mm}
		\hline\noalign{\vskip 0.6mm}
	\end{tabular}
	\par
\end{table*}

Since the high-voltage driver we employed is digital (only fixed half-wave voltage can be applied to the Pockels cells), according to the architecture of Loop memory and the birefringence of wave plates, photons transmitted through the crystal twice will be mapped out by more than 75\% probability. As is shown in Figure 9(a), we calculate the probability distributions of mapped out photons as functions of half-wave voltage, after multiple passing through the crystal from the first circle to the tenth. We can see that after transmitting through the crystal twice, photons can be mapped out with a high probability, thus we mainly consider the mapped-out pulses of the first and second rounds. Therefore, in this demonstration, only two temporal modes of AS1(AS2) and AS1'(AS2') are presented in Figure 9(b), while actually any possible probability distributions and arbitrary temporal modes with fixed time interval can be realized, with an analogue high-voltage driver.

\begin{figure*}
	\centering
	\includegraphics[width=2\columnwidth]{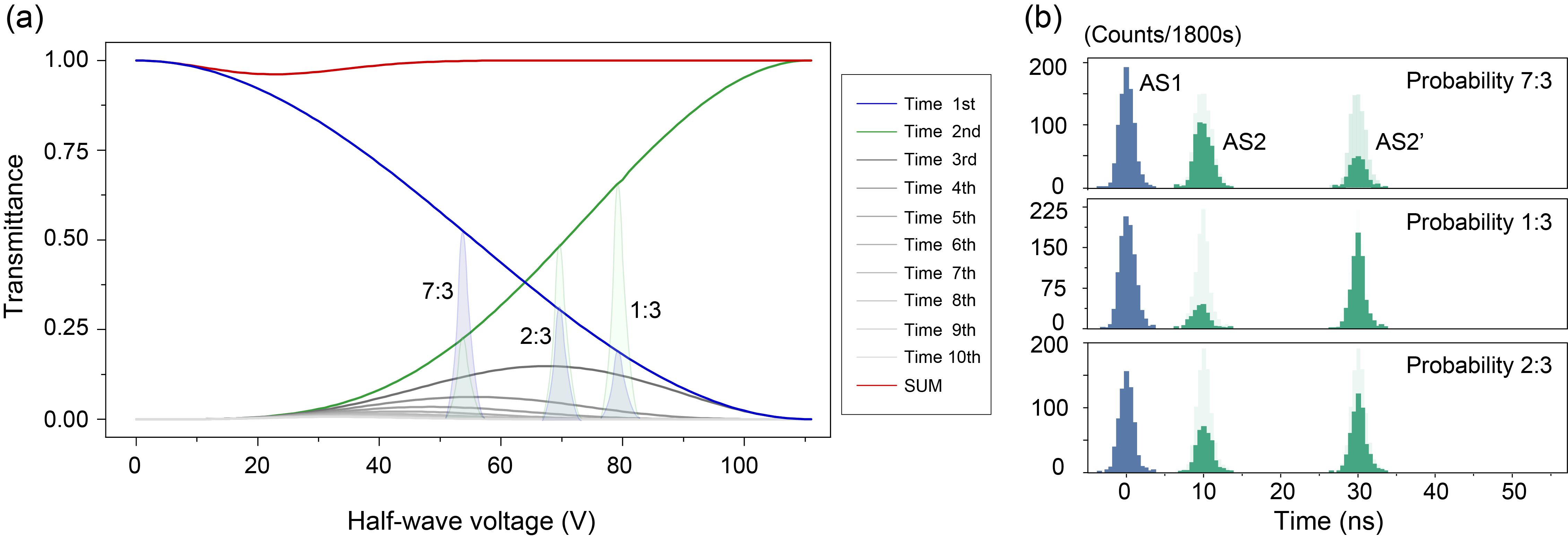}\\
	\caption{\textbf{Tunable chopping probability of a single photon in the Loop memory.} \textbf{(a)} Calculated probability distributions of the mapped-out photons as functions of half-wave voltage. Focus on the first (curve in blue) and second part (curve in green), the proportion between two pulses can be arbitrary such as 7:3, 2:3 and 1:3 marked as pulse shadows, by applying different half-wave voltages to the Pockels cells. \textbf{(b)} The measured probability distributions of the mapped-out anti-Stokes photons. The shadow pulses represent the sum of AS2 and AS2'.}
	\label{f5}
\end{figure*}

\begin{figure*}
	\centering
	\includegraphics[width=2\columnwidth]{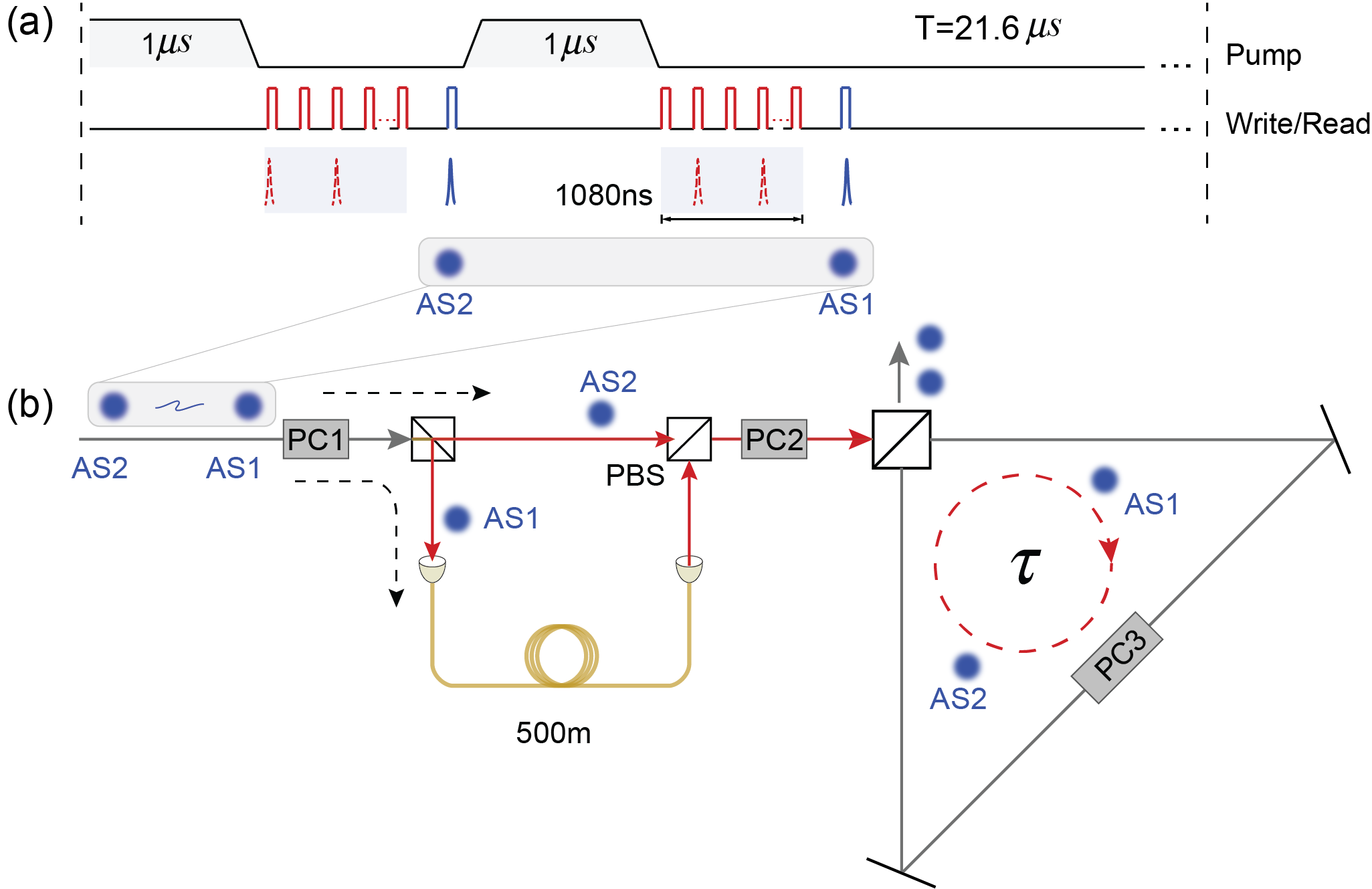}\\
	\caption{\textbf{Path selection process before photons being mapped into the Loop memory.} \textbf{(a)} Time sequences of pumping light and write/read pulses for generating a two-photon chain in FORD memory with memory-enabled feedback enhancement. The operation period of 21.6\,$\mu$s is limited by the repetition rate of PC1 and PC2, which is 50\,kHz. \textbf{(b)} Schematic diagram of path selection process. A 500-meter-long fiber serves as a low-loss optional delay line for anti-Stokes photons. PC: Pockels cell. Circulation period $\tau$ equals to 21.6\,$\mu$s.}
	\label{f6}
\end{figure*}

Note that the excitation rate of the FORD memory here is enhanced by up to 10 times, by a newly developed feedback circuit, which can further bring exponential enhancement of multi-photons synchronization. As shown in the schematic diagram Figure 7(b), we repeat the write process for 10 times in a period (21.6\,$\mu$s) until a successful Stokes photon is detected, then we halt the write process with a homemade feedback circuit based on FPGA. The enhancement of the success rate of preparing heralded single photons we obtained here depends almost linearly with the feedback times. In our new experiment, we achieve 10 times higher excitation rate of single photons, which greatly facilitates the efficient realization of storing a fully operable heralded photon chain state. 

In order to demonstrate high nonclassicality, the atomic ensemble is pumped to initialize the state each time before the write/read pulses applied, which takes 1\,$\mu$s for pumping, as is shown in Figure 10(a). Meanwhile, it takes another 1.08\,$\mu$s to repeat write pulses until a successful Stokes photon is detected. Including the optical path delay, the interval between two heralded anti-Stokes photons is more than 2.5\,$\mu$s, without considering the storage time of FORD memory. Therefore, a 500-meter-long fiber is introduced as a fixed delay line to coordinate with the Loop memory. The fixed delay of the switchable fiber is employed to mitigate the lifetime difference between FORD and Loop quantum memories, allowing versatile two-memory coordination.

To be specific, we introduce a 500-meter-long fiber and two optical switches, to coordinate with the Loop memory, as low-loss delay. As we can see in Figure 10(b), two sets of Pockels cell and polarization beam splitter serve as high-speed optical switches, which are controlled by the signals of FPGA. Pockels cells are activated when the first anti-Stokes photon arrives. The first Pockels cell switches the photon to be vertically polarized, and then the photons will be reflected into the 500-meter-long fiber. Then the second Pockels cell switches the photon back to be horizontally polarized. As for the second anti-Stokes photon, Pockels cells will not respond. Such a 500-meter-long fiber is specifically designed for long storage time scenarios. For example, in order to map a two-photon chain state into the Loop memory, the first arrived anti-Stokes photon will be stored for more than 2.5\,$\mu$s in loop to wait for the arrival of the second anti-Stokes photon, leading to significant photon loss of the first anti-Stokes photon. The long fiber here provides another path with additive latency, which is relatively low-loss, and can coordinate with the Loop memory, to demonstrate the temporal swapping of photons in a chain.

\end{document}